\documentclass[12pt, prd, a4paper, aps, preprintnumbers]{revtex4}
\pdfoutput=1
\usepackage{mathrsfs,}
\usepackage{color}
\usepackage[active]{srcltx}
\usepackage{amsmath,amsfonts,amssymb,amsthm,amstext,amscd,eucal,srcltx}
\usepackage{epsfig,graphicx,bm}
\usepackage{epstopdf, epsf}
\usepackage{dcolumn}% Align table columns on decimal point

\DeclareMathOperator{\sech}{sech}

\def\nn{\nonumber}

\newcommand{\mpl}{M_{\rm pl}}
\def\AS{{\alpha_{\vec k}^*}}
\def\Ai{{\alpha_{\vec k_i}}}

\def\Ais{{\alpha_{\vec k_i}^{*}}}
\def\Ajs{{\alpha_{\vec k_j}^{*}}}
\def\A{{\alpha_{\vec k}}}
 \def\B{{\beta_{\vec k}}}
\def\Bi{{\beta_{\vec k_i}}}

\def\Bis{{\beta_{\vec k_i}^{*}}}
\def\Bjs{{\beta_{\vec k_j}^{*}}}
 \def\BS{{\beta_{\vec k}^*}}

\def\chis{\chi_{{}_S}}

\def\phis{\varphi_{{}_S}}
\def\vare{{\varepsilon}}
\def\b1{{\beta_1}}

\def\nn{\nonumber \\ }
\newcommand{\lp}{\left(}
\newcommand{\rp}{\right)}
\newcommand{\lb}{\left[}
\newcommand{\rb}{\right]}

\newcommand{\be}{\begin{equation}}
\newcommand{\ee}{\end{equation}}
\newcommand{\beqa}{\begin{eqnarray}}
\newcommand{\eeqa}{\end{eqnarray}}

\begin{document}

%\preprint{UAB-FT-xxx}

\title{Anisotropic non-Gaussianity from Rotational Symmetry Breaking Excited Initial States}

\author{Amjad Ashoorioon}
\email{amjad.ashoorioon@bo.infn.it}
\affiliation{I.N.F.N., Sezione di Bologna, IS FLAG
viale~B.~Pichat~6/2, I-40127 Bologna, Italy}

\author{Roberto Casadio}
%\email{konstand@ifae.es}
\affiliation{Dipartimento di Fisica e Astronomia,
Alma Mater Universit\`a di Bologna,
via~Irnerio~46, 40126~Bologna, Italy}
\affiliation{I.N.F.N., Sezione di Bologna, IS FLAG
viale~B.~Pichat~6/2, I-40127 Bologna, Italy}

\author{Tomi Koivisto}
\affiliation{Nordita, KTH Royal Institute of Technology and Stockholm University,
Roslagstullsbacken 23, SE-10691 Stockholm, Sweden}

%\date{\today}

\begin{abstract}
If the initial quantum state of the primordial perturbations broke rotational invariance, that would be seen as a statistical anisotropy in the angular correlations of the cosmic microwave background radiation (CMBR) temperature fluctuations. This can be described by a general parameterisation of the initial conditions that takes into account the possible direction-dependence of both the amplitude and the phase of particle creation during inflation. The leading effect in the CMBR two-point function is typically a quadrupole modulation, whose coefficient is analytically constrained here to be $|B| \lesssim 0.06$. The CMBR three-point function then acquires enhanced non-gaussianity, especially for the local configurations. In the large occupation number limit, a distinctive prediction is a modulation of the non-gaussianity around a mean value depending on the angle that short and long wavelength modes make with the preferred direction. The maximal variations with respect to the mean value occur for the configurations which are coplanar with the preferred direction and the amplitude of the non-gaussianity increases (decreases) for the short wavelength modes aligned with (perpendicular to) the preferred direction. For a high scale model of inflation with maximally pumped up isotropic occupation and
$\epsilon\simeq 0.01$ the difference between these two configurations is about $0.27$, which could be detectable in the future. For purely anisotropic particle creation, the non-Gaussianity can be larger and its anisotropic feature very sharp. The non-gaussianity can then reach $f_{NL} \sim 30$ in the preferred direction while disappearing from the  correlations in the orthogonal plane.
\end{abstract}

\maketitle

\section{Introduction}

The inflationary paradigm is well compatible with the latest Planck 2015
data~\cite{Ade:2015lrj}.
Built upon quantum field theory and general relativity, the predictions of this
scenario also depend on symmetries that are known to hold at low energy.
However, since the energy scale of inflation could be as large as the GUT scale,
it is not unreasonable to consider that some of these symmetries were
broken during inflation.
In particular, rotational symmetry could have been broken by the anisotropic
evolution of the background, due to a contribution in the energy density of the
universe that breaks the symmetry at the classical level, or at the quantum level.
One could envisage that the effect of high-energy physics was to excite some
modes of the primordial quantum fluctuations from the Bunch-Davies (BD) vacuum
to excited states that break the rotational symmetry by picking up a preferred
direction.
These anisotropic excited states set the initial condition for the perturbations
inside the horizon that will get stretched to scales well beyond the horizon
during the exponential but still isotropic expansion of the universe, and eventually
re-enter to give rise to the cosmic microwave background radiation (CMBR).
\par
On the other hand, if inflation lasted only for a finite period, it may not have had time to 
completely smoothen and isotropise the universe at the very largest scales. 
The perturbation modes at those scales would then originate from a background 
without the symmetries of the Minkowski space, and in this case one may also
expect the initial vacuum state to break the isotropy of the BD vacuum. 
Two complementary interpretations for such a breaking are possible because the largest
scales of our universe have stayed frozen outside the horizon for longest
and thus at the same time carry the imprints of the earlier stage of the primordial
vacuum.
\par
In this work, we will consider a general parameterisation of the initial conditions
that can describe both types of scenarios.
We assume a preferred direction $\hat n$ in
momentum space~\footnote{We denote unit vectors with a hat.}, 
which generically leads to the modified scalar power spectrum
\be
\label{P-parity}
{\mathcal P}_{S}
=
{\mathcal P}_{\rm iso} \left[1+  M(\hat k)\right]
\ ,
\ee
where ${\mathcal P}_{\rm iso}$ is the standard isotropic spectrum
and 
\be
M(\hat k)
=
A \,\hat{k}\cdot\hat{n}
+
B\,(\hat{k}\cdot\hat{n})^2
+
C\,(\hat{k}\cdot\hat{n})^3
+
\ldots
\ ,
\label{Mk}
\ee
contains the dipole (quadrupole) term proportional to $A$ ($B$) as well
as higher multipoles.
The same pattern will then appear in the temperature anisotropy of the CMBR,
that is
\be
\label{DeltaT}
\Delta T(\hat k)
=
\Delta T_{\rm iso}(\hat k) \left[1+M(\hat k)\right]
\ .
\ee
\par
The coefficient $A$ and all odd higher multipoles vanish, as they cannot be accommodated in 
a scalar spectrum (except in a non-commutative model of inflation \cite{Koivisto:2010fk}), as explained in detail in the appendix~\ref{AppA}. 
A previous study considered the possibility of a - somewhat ad hoc - dipole modulation of the initial conditions in the 
position space, with the motivation of generating a hemispherical asymmetry in the 
CMBR \cite{Ashoorioon:2015pia}. In this paper we however parameterise the properties of the initial state in the momentum space,
by assigning initial conditions for each Fourier mode such that they depend only upon the wavevector of the mode.
The odd contributions then drop out in (\ref{Mk}).
\par
Higher-order multipoles could also contribute in~\eqref{P-parity} and~\eqref{DeltaT}.
For example, Ref.~\cite{Ackerman:2007nb} suggested an inflationary model that
can account for the quadrupole term in~\eqref{Mk}, and computed the effects
such modification of the power spectrum would have on the CMBR~\eqref{DeltaT}. Various other models, most containing gauge fields 
during \cite{Maleknejad:2012fw}, or after inflation \cite{Dimopoulos:2006ms}, have been suggested as the origin of a quadrupolar term in the primordial power spectrum. 
Komatsu and Kim~\cite{Kim:2013gka} tried to constrain $B$ using the Planck 2013
data~\cite{Ade:2013ydc} and found no evidence for this violation of rotational
symmetry.
After the removal of the beam asymmetry effect in the Planck 143 GHz map, they
found 
\be
\label{B-KK}
-0.03<B<0.033~(95\%~{\rm C.L. })
\ .
\ee
The Planck collaboration put model-dependent constraints on $B$
from their 2013 data~\cite{Ade:2013ydc} in the context of different anisotropic
inflationary models (see~\cite{Maleknejad:2012fw} for a handful of these models)
exploiting the effects of the quadrupole term on the three-point function.
The strongest of such constraints is
\be
-0.05<B<0.05~(95\%~{\rm C.L. }).
\ .
\ee 
Kamionkowski and Pullen \cite{Pullen:2007tu} claim that Planck can detect the quadrupole power as small as $2\%$.
In this work, we shall also obtain a crude bound on the parameter $B$
analytically from the lack of violation of statistical cosmic isotropy at high $l$'s,
which we find in agreement with~\eqref{B-KK} from~\cite{Kim:2013gka}.
The way this bound is obtained implies that we could obtain a better estimate of the parameter $B$
from improved data at higher $l$'s. 
\par
The main aim of this work is to understand the general implications of the possible
anisotropy in the initial conditions to the observable correlations of the CMBR temperature.
Besides the statistical anisotropy in the two-point function, there will 
also be signatures in the bispectrum.
The amplitude of the local non-gaussianity gets generally enhanced in
the presence of excited states.
However, with an anisotropic power spectrum, the amplitude of $f_{\rm NL}$
will also depend on the angles the modes make with the preferred direction.
For positive $B$, in a triangular configuration, when the short wavelength modes
are parallel (or antiparallel) to the preferred direction, we will get the largest
increment to the amplitude of $f_{\rm NL}^{\rm local}$.
On the other hand, if the long wavelength mode is (anti)parallel to the
preferred direction, we will get the maximal reduction from the mean value for
the $f_{\rm NL}^{\rm local}$ (the situation is reversed if the parameter $B$
is negative). 
\par
The structure of the paper is as follows.
In Section~\ref{secII}, we formulate our general parameterisation of the initial
conditions, and discuss two distinct classes of physical models which correspond
to different regions of the parameter space. In the presence of physics that
violate the symmetries of the cosmological background, the question arises if
one can still assume this background to arise as a supposed average of its contents:
we take this issue of backreaction carefully into account in Section~\ref{secII} and
go to further detail in Appendix \ref{AppC}.
We then compute the effect of the quadrupole term on the temperature 
anisotropy of the CMBR in Section~\ref{secIII}.
In particular, we find an analytic bound on the parameter $B$, which quantifies
the amount of violation of the rotational invariance, comparable with the bounds
obtained using the Fisher Matrix methods.
In Section~\ref{secIV}, we obtain the bispectrum in the anisotropic scenarios.
As expected, the local configuration is enhanced for such excited
initial states with an amplitude which is within the $1\sigma$ bound
of the Planck data.
In addition, one finds a modulation that depends on the angles
the modes make with the preferred direction.

\section{Rotational Symmetry Breaking Excited Initial States}
\label{secII}
The predictions of inflationary models for the CMBR spectrum depend
on the initial state of the quantum perturbations as well as the specific
details of the model. The standard lore is that these perturbations embark upon the Bunch-Davis
(BD) vacuum~\cite{Bunch:1978yq}, and are therefore minimum energy states
at the time they pop out of vacuum inside the horizon of an inflationary
background.Therefore it is conceivable, and in fact can be shown, that the predictions of inflation depend on the initial condition of perturbations.
\par
The equation of motion for the gauge-invariant scalar perturbation modes, $u_{\vec k}$, in an FRW background is
\begin{equation}
\label{u-eq}
u^{\prime\prime}_{\vec k}+\left(k^2-\frac{z^{\prime\prime}}{z}\right)u_{\vec k}=0
\ ,
\end{equation}
where $u_{\vec k}(\tau)$ is the spatial Fourier mode of the Mukhanov-Sasaki
variable, as defined in \cite{Mukhanov:1990me}. Prime denotes derivative with respect to the conformal time $\tau$. 
In a quasi-de-Sitter background, where the Hubble parameter $H$ is almost constant, the most general solution is given by
\begin{equation}
\label{u-sol-ds}
u_{\vec k}(\tau)
\simeq
\frac{\sqrt{\pi|\tau|}}{2}\left[\A\, H_{3/2}^{(1)}(k|\tau|)+\B\, H_{3/2}^{(2)}(k|\tau|)\right]
\ ,
\end{equation}
where $H_{3/2}^{(1)}$ and $H_{3/2}^{(2)}$ are respectively Hankel functions
of the first and second kind, which respectively behave like the positive and negative
frequency modes in the infinite past. The Bogoliubov coefficients satisfy the Wronskian constraint,
\begin{equation}
\label{Wronskian}
|\A|^2-|\B|^2=1
\ ,
\end{equation}
and the standard BD vacuum is obtained when $\A=1$ and $\B=0$. 
We allow these coefficient to depend on the direction of the excited
momenta, with a form that will depend on the excitation mechanism.
For example, if the inflaton is coupled to a vector field, it is conceivable
that these coefficients depend on the direction of momenta.
\par
The scalar power spectrum,
\begin{equation}
{\mathcal P}_{S}
=
\frac{k^{3}}{2\pi ^{2}}\left| \frac{u_{k}}{z}\right|^2_{{k/{\cal H}\rightarrow 0}}
\ ,
\label{scrpower}
\end{equation}
now turns out to be a modulation of the BD spectrum, that is
\begin{equation}
\label{power-spectrum-scalar}
{\mathcal P}_S={\mathcal P}_{\rm BD}\,\gamma_{S} ,
\ee
where
\be
\label{P-BD-gamma-S}
{\mathcal P}_{\rm BD}
=
\frac{1}{8\pi^2\epsilon}\left(\frac{H}{\mpl}\right)^2
\ ,
\qquad
\gamma_{S}=|\A-\B|^2_{{}_{k={\cal H}}}
\ .
\end{equation}
We note in particular that the power spectrum (like the bi-spectrum)
only depends on the relative phase of $\AS$ and $\BS$.
Hence, it is convenient to parameterise the isotropic contribution as
\be
\label{parametrization}
\alpha^S_{k} 
=
e^{i\varphi_{{}_S}} \cosh\chi_{{}_S} 
\ ,
\qquad
\beta^S_{k}
=
e^{-i\varphi_{{}_S}} \sinh\chi_{{}_S} 
\ ,
\ee
so that $\chi_{{}_S}\simeq\sinh^{-1}\beta_k$
and $e^{-2\chi_{{}_S}}\leq \gamma_{{}_S}\leq e^{2\chi_{{}_S}}$. In an anisotropic vacuum state, the parameters $\chis$ and $\varphi_{{}_S}$ can be direction-dependent.
\par
The most general form of~\eqref{parametrization} up to second order in
$\hat k \cdot \hat n\equiv \cos\psi_{\vec k}\equiv c_{\hat k}$ is then
\be \label{beta-expansion}
\beta_0(\hat{k})  = 
\sinh\lp \chis + \vare_2\,c_{\hat k}^2\rp
e^{-i\left(\phis + \delta_2\, c_{\hat k}^2\right)}\,.
\ee
In accordance with the Wronskian constraint~\eqref{Wronskian}, the first Bogoliubov coefficient is
\be 
\alpha_0({\hat{k}}) = \cosh\lp \chis + \vare_2\,c_{\hat k}^2\rp e^{i\left(\phis + \delta_2\, c_{\hat k}^2\right)}\,.
\ee
Now the isotropic amplitude and phase difference of the Bogoliubov coefficients is parameterised by $\chis$ and $\delta_2$,
and we also take into account the angularly varying effects, given by $\vare_2$ and $\delta_2$, respectively. We do not include odd dependence on the wave vector as explained in the appendix \ref{AppA}.  As one expects, the rotation-breaking contribution is constrained to be perturbative small, but it can dominate over a possible isotropic contribution as we shall see. 
The general expression for the quadrupole modulation (where we do not assume $|\vare_2|,|\delta_2|<1$), can be written as
\be
B = \frac{2\lp \vare_2 +\delta_2\sin 2\phis\rp\sinh 2\chis - 2\vare_2\cos 2\phis \cosh 2\chis }{\cosh 2\chis - \cos 2\phis \cosh 2\chis}\,.
\ee
This is the generic leading order anisotropic contribution. Before going into more detailed predictions, we will first consider the backreaction constraints, 
due to which it is necessary to take into account also scale dependence.
\par
For a generic initial state,  the energy and pressure density carried by the
fluctuations are of the same order,
$\delta p_{\text{non-BD}} \sim \delta \rho_{\text{non-BD}}$,
and should remain subdominant with respect to the inflaton total energy.
Their variations with time should also not hinder the slow-roll condition.
Noting that
$\delta\rho_{\text{non-BD}}'\sim \delta p_{\text{non-BD}}'\sim {\cal H}\, \delta\rho_{\text{non-BD}}$
in the leading slow-roll approximation, the latter requirement is satisfied if
\be
\label{background-backreaction}
\delta\rho_{\text{non-BD}} \ll \epsilon\,\rho_0\,,
\qquad
\delta p_{\text{non-BD}}'\ll {\cal H}\,\eta\,\epsilon\,\rho_0
\ ,
\ee
where $\epsilon$ and $\eta$ are first and second  slow-roll coefficients parameters (for definitions please see \cite{Ashoorioon:2014nta})
and the strongest of the above two constraints may be written in terms
of $\beta_k$ as
\be \label{enonbd}
\int_H^{\infty} \frac{{\rm d}^3 k}{(2\pi)^3} k |\B|^2
\ll
\epsilon\,\eta\, H^2\mpl^2\ .
\ee
We can then discuss the possible physical mechanisms of vacuum excitations and check whether they could be realised 
consistently for perturbations in the FRW background.
\subsection{High-energy new physics above $M$} 
Various effects of physics at energy scales higher than that
of inflation~\cite{Initial-data-literature}, or multi-field effects~\cite{Shiu:2011qw},
could have excited these fluctuations to a state above the Bunch-Davies 
vacuum \cite{Mukhanov:1990me}. 
\par
As was shown in~\cite{Ashoorioon:2013eia}, in the regime where
the deviation from the BD vacuum is large, $\chis \gg 1$,
in order to have maximal separation between the scale $M$ of new physics
and the inflationary Hubble parameter $H$, one is confined to
$\phis\simeq \pi /2$ and the Bogolubov coefficients are purely
imaginary~\footnote{For the inflaton potential $m^2\,\phi^2$,
$\chis \gg 1$ yields $M\simeq 21\,H$.}. The maximum separation is desirable
for the validity of the effective field theory approach.
It has also been shown that by assuming initial conditions
other than the BD vacuum both for scalar and tensor perturbation, one can decrease
the tensor-to-scalar ratio in a high-energy scale chaotic models like
$m^2\phi^2$~\cite{Ashoorioon:2013eia}
and make it compatible with the latest Planck data~\cite{Ade:2015lrj,Ade:2013ydc},
and that a large amount of running of the scalar
spectral index or a blue tensor spectral index can be induced using scale-dependent
initial condition~\cite{Ashoorioon:2014nta}. 
\par
We assume that all scales of interest are uniformly excited to an initial
state with the second Bogoliubov coefficient,
\be
\label{beta01}
\beta_{\vec k}=\theta(aM-k)\beta_0(\hat k)\,
\ee
once their physical momenta become smaller than the scale $M$
of new physics, that is $k/a(\tau)\lesssim M$.
Inevitably, modes which remain above this threshold do not get
excited.
With sufficiently high $M$ the choice~\eqref{beta01} does not lead to any extra $k$-dependence in the power
spectrum and does not change the spectral index at observable scales.
Moreover, since we have
\begin{equation}
\label{p-rho-massless-quanta}
\delta \rho_{\text{non-BD}} \sim 
\delta p_{\text{non-BD}}'/{\cal H}
\sim
|\beta_0(\hat k) |^2 M^4
\ ,
\end{equation}
one obtains the upper bound 
\begin{eqnarray}
\label{beta-scalar-backreaction}
|\beta_0 (\hat k)|
\lesssim
\sqrt{\epsilon\,\eta}\,\frac{H M_{\rm Pl}}{M^2}
\sim
\epsilon\,\frac{H M_{\rm Pl}}{M^2}
\ .
\end{eqnarray}
As discussed in~\cite{Ashoorioon:2013eia} and will be reviewed briefly below,
this does not mean $|\beta_0(\hat k)|$ is necessarily very small.
Larger values of $|\beta_0 (\hat k)|$ can in fact be compensated by a smaller Hubble
parameter $H$ in order to match the normalization of density perturbations
with the data.

In the case of imaginary coefficients, $\phis=\phis^{(S)}=\pi/2$, which we denote with the upper index $S$ for clarity, we obtain that
\beqa
\gamma_S^{(S)} & = & e^{2\chis} + 2\vare_2 e^{2\chis} c_{\hat k}^2 \nonumber \\
& + & 2\lp \vare_2^2 e^{2\chis} -\delta_2^2\sinh 2\chis\rp  c_{\hat k}^4 + \dots\,,
\eeqa
and thus the quadrupole is simply $B^{(S)}=2\vare_2$. As the power spectrum is amplified by ${\mathcal P}_S^{(S)}=e^{2\chis}{\mathcal P}_{BD}$, it is obvious that when matching the result with the observed amplitude of the primordial fluctuations one is forced to lower the inflationary scale exponentially with increasing $\chis$. In this case, the observational constraint~\eqref{B-KK} yields
\be
\label{vare2}
-0.015 <\vare_2<0.0165~(95\%~{\rm C.L.})\,,
\ee
but leaves $\delta_2$ as a free parameter.

\subsection{Large-scale asymmetry below $M$}

It is conventional to assume that inflation lasted long enough so that at each relevant cosmological scale, the perturbation modes had been safely inside the BD vacuum region in the past. However, if inflation went on only for approximately the minimal period that is needed to generate a homogeneous universe of the present proportions, that is about 60 e-folds, then the largest observable scales originate effectively from a background that inflation did not have time to homogenise. As such modes live in a background geometry that does not have the symmetries of the Minkowski space, there seems no reason to expect that their quantum state should somehow have settled to the Minkowski vacuum. 

Thus, even in the context of the most minimal single-field inflation models, it can naturally occur that perturbations at the largest relevant scales had frozen outside the horizon in non-BD state, if we just consider that inflation started from a generic initial state and lasted only for a finite period. One may argue, that due to the accumulation of quantum backreaction, there are generic restrictions on the duration of inflation \cite{Koivisto:2010pj}. Explicit boundary conditions have been also formulated for the start of inflation by quantum tunneling from e.g. Kasner spacetime \cite{Kim:2011pt,Blanco-Pillado:2015dfa}, in which case the largest scales of our universe would indeed carry imprints from the anisotropies of the primordial vacuum. 

In addition, inflation itself could be (slightly) anisotropic. There are models where inflation has some small anisotropic ''hair'' generated by vector field dynamics, that can be present only for finite period without backreacting too much on the average isotropy \cite{Maleknejad:2012fw,Dimopoulos:2006ms,Koivisto:2014gia,Watanabe:2009ct}. In such models one may well consider that the vacuum state should reflect the non-minimal vector field couplings and the anisotropy of the background by breaking rotational invariance, see Ref. \cite{Chen:2013tna,Emami:2014tpa,Chen:2014vja}.  

In scenarios such as discussed above, the cut-off scale $M$ can be seen as an infrared scale, below which the rotational symmetry is broken. Basically there is a length scale that marks distances above which the anisotropies do not average out, and the assumption of the maximally symmetric vacuum breaks down. One can then regard that the particle creation stems exclusively from the breaking of symmetry: in the case of a preferred direction puncturing the sphere of SO(3), we would have purely homogeneous deviation from BD vacuum, and in such a way that it vanishes on the plane of the residual symmetry, wherein the modes would have the standard initial conditions. The corresponding parameter region is $\chis \approx 0$.

In this case of purely anisotropic deviation from the Bunch-Davis vacuum, $\chis = 0$, we denote $\gamma_S$ by the upper index $\gamma_S^{(0)}$, and similarly for other quantities.
We obtain the $\gamma_S$ from the relation~\eqref{P-BD-gamma-S} as,
\beqa
\gamma_S^{(0)} & = & 1 - 2\vare_2\cos 2\phis c_{\hat k}^2 
 +  2\vare_2\lp \vare_2 + 2 \delta_2\sin 2\phis\rp c_{\hat k}^4 \nonumber 
 \\ & + & 4 \vare_2\cos 2\phis\lp \delta_2^2 - \frac{1}{3}\vare_2^2\rp c_{\hat k}^6 + \dots \,.
\eeqa
Thus, by writing the power spectrum as in \eqref{P-parity}, the leading order coefficients
\be\label{B-anisotropic}
A^{(0)} = 0\,, \quad B^{(0)} = -2\vare_2\cos 2\phis\,, \quad C^{(0)}=0\,,
\ee 
all vanish if the phase difference is a right angle, $2\phis=\pi/2$. The higher order coefficients will not vanish in general, but they are suppressed by the corresponding powers of $\vare_2$ and $\delta_2$. See the Appendix \ref{AppA} for more details.

\section{Microwave Background}

In the following, we will analytically find a bound on $B$
comparable to that found in~\cite{Kim:2013gka}.

\label{secIII}
We would like to know how the modified power spectrum~\eqref{P-parity}
affects the prediction for the CMBR temperature fluctuations
up to the quadrupole term $B$.
The anisotropy in $\Delta T/T$ along the direction of the unit vector
$\hat e$ is related to the primordial fluctuations by
\begin{equation}
\label{def2}
{\Delta T \over T}({\hat e})
=
\int{\rm d}{\vec k}\,
\sum_{l}\left({2l+1 \over 4 \pi}\right)
(-i)^l\, P_l({\hat {k}}\cdot {\hat e})\, \mathcal{R}({\vec k})\, \Theta_l(k)
\ ,
\end{equation}
where $P_l$ is the Legendre polynomial of order $l$, $\Theta_l(k)$
is a function of the magnitude $k=|\vec k|$ that incorporates, for example,
the effects of the transfer function~\footnote{The transfer function is assumed
to depend on $k$ since the dynamics after the inflationary era is presumed
to be rotationally invariant.}, and $\mathcal{R}({\vec k})=-u_{\vec k}/z$ at the
end of inflation. 
The CMBR multipole moments are then defined by
\begin{equation}
\label{def1}
a_{l m}
=
\int {\rm d} \Omega_{\hat e} \,[Y_l^m({\hat e})]^*{\Delta T \over T}({\hat e})
\ .
\end{equation}
\par
Since we are interested in computing the expectation values
$\langle a_{lm}\,a_{l'm'}^* \rangle$
to leading order in the small quantity $\vare_2$, we write 
\begin{equation}
\label{correlationmatrix}
\langle a_{lm}\,a_{l'm'}^* \rangle
=
\langle a_{lm}\,a_{l'm'}^* \rangle_0+ \Delta(lm;l'm')
\  ,
\end{equation}
where the usual isotropic part is
\begin{equation}
\langle a_{lm}\,a_{l'm'}^* \rangle_0
=
\delta_{l l'}\,\delta_{m m'}
\int_0^{\infty}\frac{ {\rm d}k}{k}\,  \mathcal{P}_{\rm iso}(k)\,\Theta^2_l(k)
\ .
\end{equation}
Like in~\cite{Ackerman:2007nb}, we work with the ``spherical" components
of the preferred unit vector ${\hat n}$,
\be
n_+=-\left( {n_x-i\,n_y \over \sqrt{2}} \right)\ ,
\
n_-=\left( {n_x+i\,n_y \over \sqrt{2}} \right)
\ ,
\
n_0=n_z
\ ,
\ee
which satisfy $n_0^2-2n_+\,n_-=1$, and
\begin{equation}
P_l({\hat k}\cdot{\hat n})
=
{4 \pi \over 2l+1}
\sum_{m=-l}^{l}
Y_l^m({\hat n}) \,[Y_l^m({\hat k})]^*
\ ,
\end{equation}
Exploiting the identity, one finds
\beqa
\label{Delta}
&&\Delta(lm;l'm')
=
(-i)^{l-l'} \times
\\ 
&&
\int_0^{\infty} \frac{{\rm d}k}{k} \,\mathcal{P}_{\rm iso}(k)
B\,\xi_{lm;l'm'}^{(2)}\Theta_l(k)\,\Theta_{l'}(k)
\ , 
\nonumber
\eeqa
where
%\beqa
%\xi_{lm;l'm'}^{(1)}
%&=&
%{\left(4 \pi \over 3\right)^{1/2}}
%\int {\rm d}\Omega_{\hat k}
%[Y_l^m({\hat k})]^* \,Y_{l'}^{m'}(\hat{k})
%\nonumber
%\\
%&&
%\times\left[n_+\,Y_1^1(\hat{k})+n_-\,Y_1^{-1}(\hat{k})+n_0\,Y_1^0(\hat{k}) \right]
%\eeqa
%and 
\beqa
\xi_{lm;l'm'}^{(2)}
&=&
{4 \pi \over 3}\int {\rm d}\Omega_{\hat k}\,
[Y_l^m(\hat{ k})]^*\, Y_{l'}^{m'}(\hat{ k})
\nonumber 
\\
&&
\times
\left[n_+\,Y_1^1(\hat{k})+n_-\,Y_1^{-1}(\hat{k})+n_0\,Y_1^0(\hat{k}) \right]^2
\,,
\quad
\eeqa
and we used the phase convention for the spherical harmonics from \cite{Arfken}.
%Also $\xi_{lm;l'm'}^{(1)}$ and $\xi_{lm;l'm'}^{(2)}$ are respectively the coefficients
%proportional to $n_i$ and $n_i\,n_j$.
\par
The integral~\eqref{Delta} contains information about the power spectrum and the
transfer function, as well as the scale-dependence of the preferred-direction effect,
whereas the constants $\xi_{lm;l'm'}^{(2)}$ are purely
geometrical and can be conveniently decomposed as 
%\be\label{ximinus1}
%\xi_{lm;l'm'}^{(1)}
%=
%n_+\,\xi_{lm;l'm'}^{+(1)}+n_-\,\xi_{lm;l'm'}^{-(1)}+n_+\,n_0\,\xi_{lm;l'm'}^{0(1)}
%\ee
%and 
\beqa
\label{xiexpansion}
\xi_{lm;l'm'}^{(2)}
&=&
n_+^2\,\xi_{lm;l'm'}^{++(2)}
+n_-^2\,\xi_{lm;l'm'}^{--(2)}
\nonumber
\\
&&
+2\,n_+\,n_-\,\xi_{lm;l'm'}^{+-(2)}
+2\,n_+\,n_0\,\xi_{lm;l'm'}^{+0(2)}
\nonumber
\\
&&
+2\,n_-\,n_0\,\xi_{lm;l'm'}^{-0(2)}+n_0^2\,\xi_{lm;l'm'}^{00(2)}
\ .
\eeqa
%The coefficients linear in the $n_i $ are~\cite{Axelsson:2011gt,Koivisto:2010fk}
%\begin{eqnarray}
%\xi_{lm;l'm'}^{0(1)}
%&=&
%\delta_{m,m'}
%\left[\sqrt{\frac{(l'-m'+1)(l'+m'+1)}{(2l'+1)(2l'+3)}}\,
%\delta_{l,l'+1} 
%\right.
%\nonumber
%\\
%&&
%\qquad
%+\left.
%\sqrt{\frac{(l'-m')(l'+m')}{(2l'-1)(2l'+1)}}\,\delta_{l,l'-1}\right]
%\end{eqnarray}
%\begin{eqnarray}
%\xi_{lm;l'm'}^{+(1)}&=&\delta_{m,m'+1}\left[\sqrt{\frac{(l'+m'+1)(l'+m'+2)}{2(2l'+1)(2l'+3)}}\,\delta_{l,l'+1}\right.\nonumber\\&&\qquad-\left.\sqrt{\frac{(l'-m')(l'-m'-1)}{2(2l'-1)(2l'+1)}}\,\delta_{l,l'-1}\right]
%\end{eqnarray}
%\begin{eqnarray}
%\label{ximinus1}\xi_{lm;l'm'}^{-(1)}&=&\delta_{m,m'-1}\left[\sqrt{\frac{(l'-m'+1)(l'-m'+2)}{2(2l'+1)(2l'+3)}}\,\delta_{l,l'+1}\right.\nonumber\\&&\qquad-\left.\sqrt{\frac{(l'+m')(l'+m'-1)}{2(2l'-1)(2l'+1)}}\delta_{l,l'-1}\right]\ .
%\end{eqnarray}
These coefficients were calculated in~\cite{Ackerman:2007nb}
and are shown in Appendix~\ref{AppB}.
%Note that the coefficients $\xi_{lm;l'm'}^{(1)}$ and  $\xi_{lm;l'm'}^{(2)}$ respectively
%correlate the $a_{lm}$ with $a_{l\pm0\{\mathrm{or}~1\},m\pm0\{\mathrm{or}~1\}}$
%and $a_{l\pm2\{1~ \mathrm{or} ~0\},m\pm2\{1 ~\mathrm{or} ~0\}}$.
\par
%Eq.~\eqref{xiexpansion} expresses the geometrical part
%of the perturbation~\eqref{Delta}.
Assuming that the breaking of rotational symmetry is scale-invariant, and
thus $B(k)=B_*$, and defining polar coordinates
$\theta_*$ and $\phi_*$ for the preferred direction, 
\begin{equation}
n_x=\sin\theta_*\,\cos\phi_*
\ , \
n_y=\sin\theta_*\,\sin\phi_*
\ , \
n_z=\cos\theta_*
\ ,
\end{equation}
these expressions can be used to constrain the three parameters $(B_*, \theta_*, \phi_*)$
observationally.
When $B(k)=B_*$, for $l=l'$ and $m=m'$, the expressions simplify as the dependence
on the power spectrum for the terms that violate rotational invariance $\Delta(lm;lm)$
is the same as the rotationally-invariant part $\langle a_{lm} a_{lm}^* \rangle_0$.
We can then find a simple expression for their ratio,
\beqa
\label{wow}
&&\frac{\Delta(lm;lm)}{\langle a_{lm}\,a_{lm}^* \rangle_0}
\\
&&
=
{B_* \over 2}\left[\sin^2\theta_*+
(3\,{\cos}^2\theta_*-1)\,{2l^2+2l-2m^2-1 \over (2l-1)(2l+3)}\right]
\ .
\nonumber 
\eeqa
%In this case the coefficients $\xi_{lm;l'm'}^{(1)}=0$, and the only contribution
%comes from the $\xi_{lm;l'm'}^{(2)}$.
We note the $a_{lm}$'s are independent random variables, and we have
\be
\label{Cl0}
\langle a_{lm}\,a_{lm}^* \rangle_0
=
C^0_{l}\, \delta_{ll'}\, \delta_{mm'}
\ ,
\ee
where we introduced
\beqa
C_l
&\equiv&
\frac{1}{2l+1}\sum_{m=-l}^{m=1} a_{lm}\, a_{lm}^*
\nonumber
\\
&=&
C_l^0+\frac{1}{2l+1}  \sum_{m=-l}^{m=l}\Delta(lm;lm)
\ .
\eeqa
From~\eqref{wow} and \eqref{Cl0}, we can calculate 
\beqa\label{dCloverCl}
\frac{\Delta C_l}{C_l}
&\simeq&
\frac{\Delta C_l}{C_l^0}
%=\frac{1}{(2l+1)C_l^0} \sum_{m=-l}^{m=l}\Delta(lm;lm)
%\nonumber
\\
&=&
\frac{ B_*}{2}
{\left[\sin^2\theta_* +(3\,\cos^2 \theta_*-1)\,
\frac{(2l+1)(2l-3)}{3(2l-1)(2l+3)}\right]}
\ ,
\nonumber
\eeqa
which we note can become negative for some range
of $\theta_*$.
\par
Noting we can allow for the statistical uncertainty 
\be
\frac{\Delta C_l}{C_l}
=
\sqrt{\frac{2}{2l+1}}
\ ,
\ee
and requiring the observational constraint that the uncertainty in
${\Delta C_l}/{C_l}$ be smaller than the statistical uncertainty allowed
for a given $l$ up to $l\simeq2500$, which is the maximum $l$
probed by Planck, one can put an upper bound on $B_*$, for a given
$\theta_*$.
For high $l$'s, the upper bound on $\vare_2$ is not sensitive to
$\theta_*$.
For $l=2500$, 
\be
|B_*|\lesssim 0.06
\ ,
\ee
which is twice as large of the bound Kim and Komatsu found
using the statistical methods~\cite{Kim:2013gka} and almost
equal to the bound the Planck collaboration found using the
data on the three-point function~\cite{Ade:2013ydc}.
As we mentioned in the Introduction, a better knowledge of
the spectrum at higher $l$'s would allow for a more accurate
estimate of $B_*$ and the quadrupole correction.

Another intriguing point about \eqref{dCloverCl} is that for $0\leq\theta_*\leq\pi/6$ and  $5\pi/6\leq\theta_*\leq\pi$, $\Delta C_1/C_1$ will have a sign different from the corresponding term for other $l$'s. For maximum positive $B_*$ that satisfies the observational constraint \eqref{B-KK}, $\Delta C_1/C_1$ is about $-0.006$ which suggests that the power spectrum with quadrupole correction tends to suppress the dipole multipole coefficient, $C_1$. Of course this amount of suppression is too small to account for the suppression seen at low $l$-multipoles \cite{Ade:2015lrj}.
\section{Bispectrum}
\label{secIV}
Let us now calculate the three-point function for the above
direction-dependent excited states to see how they modify
the bispectrum.
One can first determine the Wightman function from the
solutions~\eqref{u-sol-ds},
 \be
 \label{Wightman}
 G_{k}^{>}(\tau,\tau')
 \equiv
 \frac{H^2}{{\dot{\phi}}^2} \frac{u_k(\tau)}{a(\tau)}\frac{u_k^{\ast}(\tau')}{a(\tau')}
 \ ,
 \ee
and the three-point function is then determined from the Wightman function
as~\cite{Maldacena:2002vr}
\beqa
\label{zeta3}
\langle \zeta_{\vec k_1}\zeta_{\vec k_2}\zeta_{\vec k_3}\rangle
&=&
-i\,(2\pi)^3 \delta^3\!\left(\sum_{i=1}^3 \vec k_i \right) \left(\frac{\dot{\phi}}{H}\right)^4 
\frac{H}{M_P^{2}}
\nonumber 
\\
&&
\times
 \int_{\tau_{0}}^0
 \frac{{\rm d} \tau}{k_3^2} 
\left[a(\tau) \,\partial_\tau G_{\vec k_1}^{>}(0,\tau)\right]
\nonumber
\\
&&
\times
\left[a(\tau)\, \partial_\tau G_{\vec k_2}^{>}(0,\tau)\right]
\left[a(\tau)\, \partial_\tau G_{\vec k_3}^{>}(0,\tau)\right]
\nonumber
\\
&&
+{\rm permutations}+{\rm c.c.}
\ .
\eeqa
The bispectrum then takes the form 
\beqa
\langle \zeta_{\vec k_1}\zeta_{\vec k_2}\zeta_{\vec k_3}\rangle
&=&
(2\pi)^3 \delta^3\!\left(\sum_{i=1}^{3} \vec k_i \right)
\frac{2 H^6 \,\sum\limits_{i>j} k_i^2\, k_j^2}{\dot{\phi}^2 M_P^{2}\,\prod\limits_{i=1}^3 (2k_i^3 )}
\\
&&
\times
\left[\mathscr{A}\, \frac{1-\cos(k_t \eta_0)}{k_t}
+ \mathscr{B}\, \frac{\sin(k_t\eta_0)}{k_t}
\right.
\nn
&&
\left.
+\sum\limits_{j=1}^{3} \mathscr{C}_j \frac{1-\cos(\tilde{k}_j\eta_0)}{\tilde{k}_j}
+ \sum\limits_{j=1}^{3} \mathscr{D}_j \frac{\sin(\tilde{k}_j\eta_0)}{\tilde{k}_j} \right]
\nonumber
\eeqa
where $k_t=k_1+k_2+k_3$ and $\tilde{k}_j=k_t-2 k_j$.
Terms proportional to $\mathscr{C}_j$ and $\mathscr{D}_j$ are respectively the ones
that can lead to enhancement in the local configuration,
$k_1\simeq k_2\gg k_3$~\cite{Agullo:2010ws}, or flattened (folded) configuration,
$k_1+k_2\simeq k_3$~\cite{Chen:2006nt}. 
The above coefficients are given by
\beqa
\mathscr{A}
&=&
\prod (\Ai-\Bi) \left(\prod \Ais+\prod\Bis\right)+{\rm c.c.}
\nn
\mathscr{B}
&=&
i\prod (\Ai-\Bi) \left(\prod\Bis-\prod \Ais\right)+{\rm c.c.}
\\
\mathscr{C}_j
&=&
\prod (\Bi-\Ai)
\left(\frac{\Bjs}{\Ajs} \prod \Ais +\frac{\Ajs}{\Bjs} \prod\Bis \right)
+{\rm c.c.}
\nn
\mathscr{D}_j
&=&
i\prod (\Ai-\Bi)
\left(\frac{\Bjs}{\Ajs} \prod \Ais -\frac{\Ajs}{\Bjs} \prod\Bis \right)
+{\rm c.c.}
\nonumber
\eeqa
\par
The enhancement of the flattened configuration is however lost in slow-roll
inflation after the projection of the bispectrum shape on the 2-dimensional
CMBR surface~\cite{Holman:2007na}.
Besides, for the large deviations from the BD vacuum, with
$\chis\gg 1$ and $\phi\simeq \pi/2$, the enhancement factor is exactly
equal to zero.
Thus we focus on the local configuration enhancement for $k_1 \simeq k_2\gg k_3$.
In this regime, the three-point function becomes
\beqa
\langle \zeta_{\vec k_1}\zeta_{\vec k_2}\zeta_{\vec k_3}\rangle
\simeq
-(2\pi)^3 \delta^3\!\left(\sum_{i=1}^{3} \vec k_i \right)
\frac{2\,H^8\, \epsilon}{\dot{\phi}^2 \prod\limits_{i=1}^3 k_i^3}
\frac{\sum\limits_{i>j}^{} k_i^2 k_j^2}{k_3}\mathcal{C}
\ ,
\qquad
%\nn
\eeqa
where 
\beqa
\mathcal{C}
&=&
\Re
\left\{\prod\limits_{i=1}^3 (\Ai-\Bi)
\left[\prod\limits_{i=1}^3\Ais\left(\frac{\beta_{\vec k_1}^*}{\alpha_{\vec k_1}^*}+\frac{\beta_{\vec k_2}^*}{\alpha_{\vec k_2}^*}\right)
\right.
\right.
\nn
&&
\left.
\left.
\qquad
+\prod\limits_{i=1}^3\Bis\left(\frac{\alpha_{\vec k_1}^*}{\beta_{\vec k_1}^*}+\frac{\alpha_{\vec k_2}^*}{\beta_{\vec k_2}^*}\right)
\right]
\right\}
\\
&=&
\Re\left[(\alpha_{\vec k_3}^*+\beta_{\vec k_3}^*) (\alpha_{\vec k_2}^*\beta_{\vec k_1}^*+\alpha_{\vec k_1}^*\beta_{\vec k_2}^*)\prod\limits_{i=1}^3 (\Ai-\Bi)\right]
\ .
\nonumber
  \eeqa
One can then employ the definition
\be
f_{\rm NL}
\equiv
-\frac{5}{6}\frac{\delta \langle \zeta_{\vec k_1}\zeta_{\vec k_2}\zeta_{\vec k_3}\rangle}
{\sum\limits_{i>j}  \langle \zeta_{\vec k_i}\zeta_{\vec k_i}\rangle \langle \zeta_{\vec k_j}\zeta_{\vec k_j}\rangle}
\ ,
\ee
and obtain
\be
f_{\rm NL}=-\frac{20}{3}\, \epsilon\,
\frac{k_1}{k_3}\frac{\mathcal{C}}{\gamma_S(\vec k_3) [\gamma_S(\vec k_1)+\gamma_S(\vec k_2)]}\,.
\ee
Below the formula is evaluated in the two limiting cases we consider.

\subsection{Maximally occupied vacuum}

We will first consier the limit $\chis\gg1$, which requires $\phis\simeq \pi/2$ for the validity of the effective field theory.
Assuming $\vare_1=\delta_1=0$, and expanding $ \delta f_{\rm NL}$ in terms of $\vare_2$ and
$\delta_2$ up to first order,
\be
f_{\rm NL}
\simeq
f_{\rm NL}^{0}+ f_{\rm NL}^{\vare_2}\vare_2+ f_{\rm NL}^{\delta_2}\delta_2
\ ,
\ee
we have
\beqa
f_{\rm NL}^{0}
&\simeq&
\frac{5\epsilon}{3} \frac{k_1}{k_3}\,,
\\
f_{\rm NL}^{\vare_2}
&\simeq&
\frac{5\epsilon}{3} \frac{k_1}{k_3}\left[c^2_{\vec k_1}+c^2_{\vec k_2}-2\,c^2_{\vec k_3}\right]\,,
\\
f_{\rm NL}^{\delta_2}
&\simeq&
0
\ ,
\eeqa
in the limit $\chis\gg1$ and $\phis\simeq \pi/2$, required for the validity of the effective field theory.
Since $c_{\vec k_i}\equiv\cos\psi_{\vec k_i}=\hat k_i\cdot\hat n$, the amplitude of the bispectrum
depends on the angles that three different momenta make with the preferred direction.
One should also note that, in the same limit, both the power spectrum and bispectrum do not
depend on $\delta_2$.
\par
The $ f_{\rm NL}^{0}$, which gives the dominant contribution to the bispectrum,
is however independent of the the angles.
We take the largest scale at which the cosmic variance is negligible to correspond
to $l=10$ and the smallest one to be the largest $l$ probed by the Planck experiment,
$l\simeq 2500$.
If one assumed that a large field model of inflation like $m^2\phi^2$ is made consistent with the
lack of B-mode observation, choosing the proper initial condition for the tensor
perturbations~\cite{Ashoorioon:2013eia}, so that $\epsilon\simeq 0.01$ is allowed,
one would obtain
 \be
f_{\rm NL}^{0}\simeq 4.17
\ .
\ee
This is still within the $2\sigma$ bound for local non-gaussianity in the Planck 2015 data \cite{Ade:2015ava}.

On the other hand, we could assume that tensor perturbations originate from the same excited
initial states as the scalar perturbations and use the unmodified consistency relation,
$r=16\,\epsilon$, and the current bound on the tensor-to-scalar ratio, $r<0.11$ ($95\%$ C.L.),
to constrain $\epsilon$.
We would then find the angle-independent part of the non-gaussianity is
\be
f_{\rm NL}^{0}\simeq 2.86
\ .
\ee
\par
At higher orders, the excited anisotropic initial condition induces a directional dependence
in the bispectrum at the first order correction.
Of course, the angles $\psi_{\vec k_i}$'s are not independent.
Let us first focus on the general case in which the preferred direction is not necessarily
coplanar with the triangular configuration.
%, please see figure \ref{3d-configuration}.
Since $k_1=k_2$ and $k_3\ll k_1$, the vectors $\vec k_1$ and $\vec k_2$ are almost
anti-collinear
and thus  
\be
\label{psi1-psi2}
\psi_{\vec k_2}\approx \psi_{\vec k_1}+\pi
\ .
\ee
The angles $\psi_{\vec k_i}$ can vary in the interval
\be
\theta\lesssim \psi_{\vec k_i}\lesssim \pi-\theta
\ ,
\ee
where $\theta$ is the acute angle the preferred direction makes with the plane of the triangle.
Using simple geometry, it can be shown that in the limit $k_3\ll k_1$, 
\be
\label{psi1-psi3-theta}
\cos^2 \psi_{\vec k_1}+\cos^2  \psi_{\vec k_3}\simeq\cos^2\theta
\ . 
\ee
Using this relation and~\eqref{psi1-psi2}, one obtains 
\be
f_{\rm NL}
=\frac{5}{3}\,\epsilon\,\frac{k_1}{k_3}
\left[1+\vare_2 \left(4 \cos^2 \psi_{\vec k_1}-2\cos^2\theta\right)\right]
\ .
\ee
\begin{figure}[t]
\includegraphics[angle=0, scale=0.33]{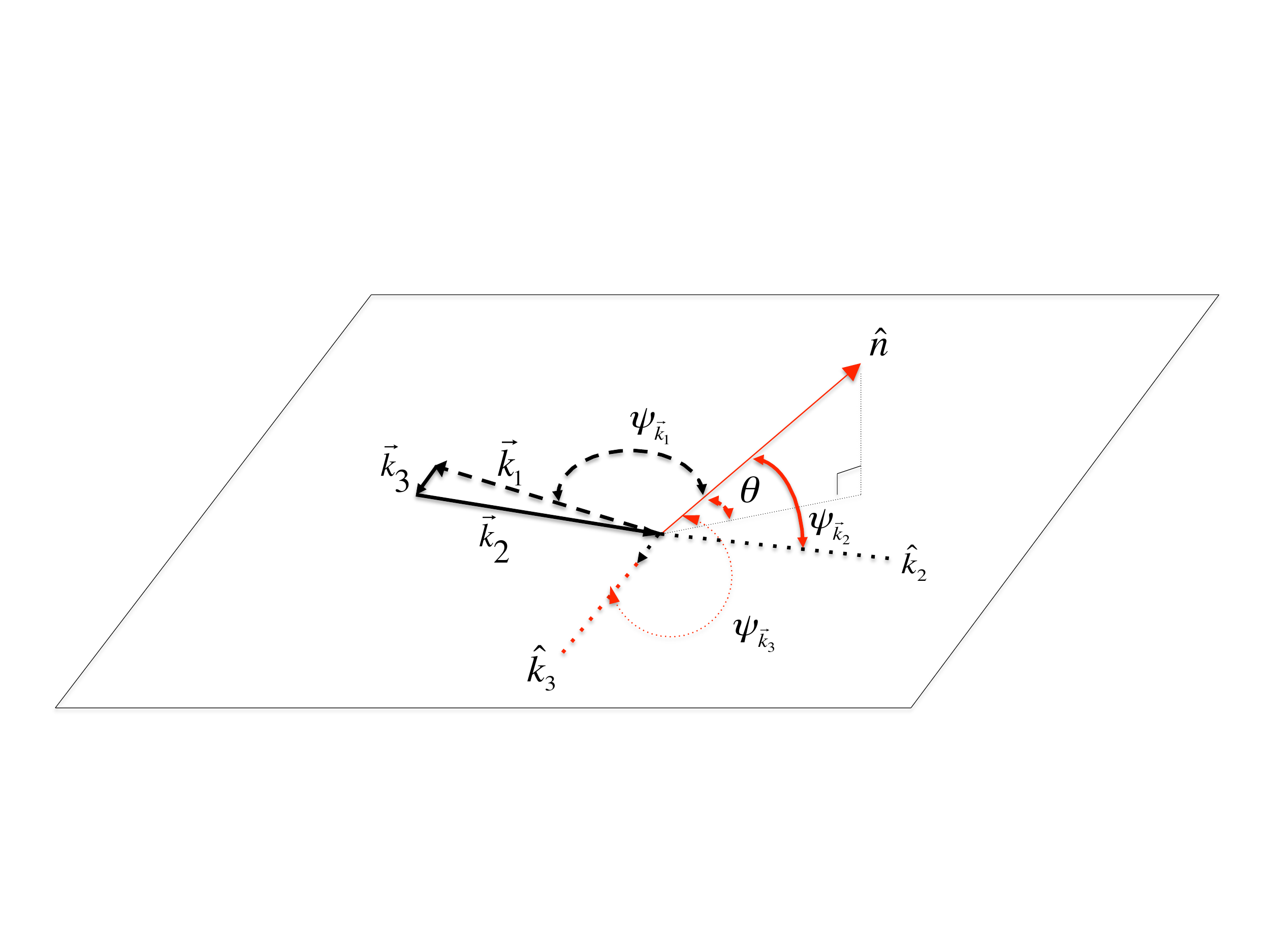}
\caption{The setup when the preferred direction, $\hat n$, makes an angle of $\theta$ with the plane of the triangular configuration.}
\label{3d-configuration}
\end{figure}

For a given $\theta$, the maximum enhancement with respect to the first
order result, $f_{\rm NL}^{0}$, is given when $\psi_{\vec k_1}=\theta$.
The minimum of $f_{\rm NL}$ would occur for a local configuration that
has $\psi_{\vec k_1}={\pi}/{2}$.
This variation between the minimum and maximum obviously enhances
for the local configuration that is coplanar with the preferred direction
$\hat n$.
The maximum and minimum for the largest positive $\vare_2$ allowed
from the data, Eq.~\eqref{vare2}, and an inflationary model with
$\epsilon\simeq 0.01$ are respectively
\be
f_{\rm NL}^{\rm max}
\simeq
4.3
\ee
and
\be
f_{\rm NL}^{\rm min}
\simeq
4.03
\ .
\ee
The maximum would occur when the largest wavenumbers are parallel
(or antiparallel) to $\hat n$.
The minimum would occur when the small wavenumber is parallel (or antiparallel)
to the preferred direction.
The difference between the values of non-gaussianity for these two configurations
is $\Delta f_{\rm NL}\simeq 0.27$ which can be used to constrain the model.
For the maximum value of $\epsilon$, one would obtain from the unmodified
consistency relation, the maximum and minimum values are
$f_{\rm NL}^{\rm max}\simeq 2.96$ and $f_{\rm NL}^{\rm min}\simeq 2.78$.
The non-gaussianity parameter, $ f_{\rm NL}$ takes intermediate values between
$f_{\rm NL}^{\rm min}$ and  $f_{\rm NL}^{\rm max}$ depending on the angle the
largest wavenumber makes with the preferred direction.
Above, we used the approximation $k_3\ll k_1$.
One can compute the corrections due to the finiteness of
$g\equiv k_3/\vec k_1$ and notice that the relative corrections
are of order $g^2$.
For $g\simeq 4\times 10^{-3}$, the relative change in non-gaussianity will be 
$\mathcal{O}(10^{-6})$.
The absolute change in the values of non-gaussianity with respect to the previous
case will be a minute $\vare \,\epsilon\, g= \mathcal{O}(10^{-6})$.  
\par
Phenomenology of models that predict a non-trivial structure in the bispectrum,
which depends on the angle between the short and long modes has been
studied~\cite{bisp}.
In our case, the modulation of the bispectrum in terms of the polar angles,
that the modes makes with the preferred direction, can be used to distinguish
this scenario.
For local configurations that are coplanar with the preferred direction,
the configurations in which the large wavenumber are parallel or antiparallel
to the preferred direction has the maximum non-gaussianity, whereas the ones
in which these modes are perpendicular to the preferred direction leads to less
amount of non-gaussianity.
The difference between non-gaussianities of these two configurations is about
$0.27$ for the largest value of $\vare_2$ (which quantifies the level of rotational
symmetry breaking) allowed  in the initial state.
This could be  which could be used to distinguish the model from other scenarios
that considers the breaking of rotational symmetry during the inflation.
\subsection{Purely anisotropic modulation}

\begin{figure}[t]
\includegraphics[angle=0, scale=0.6]{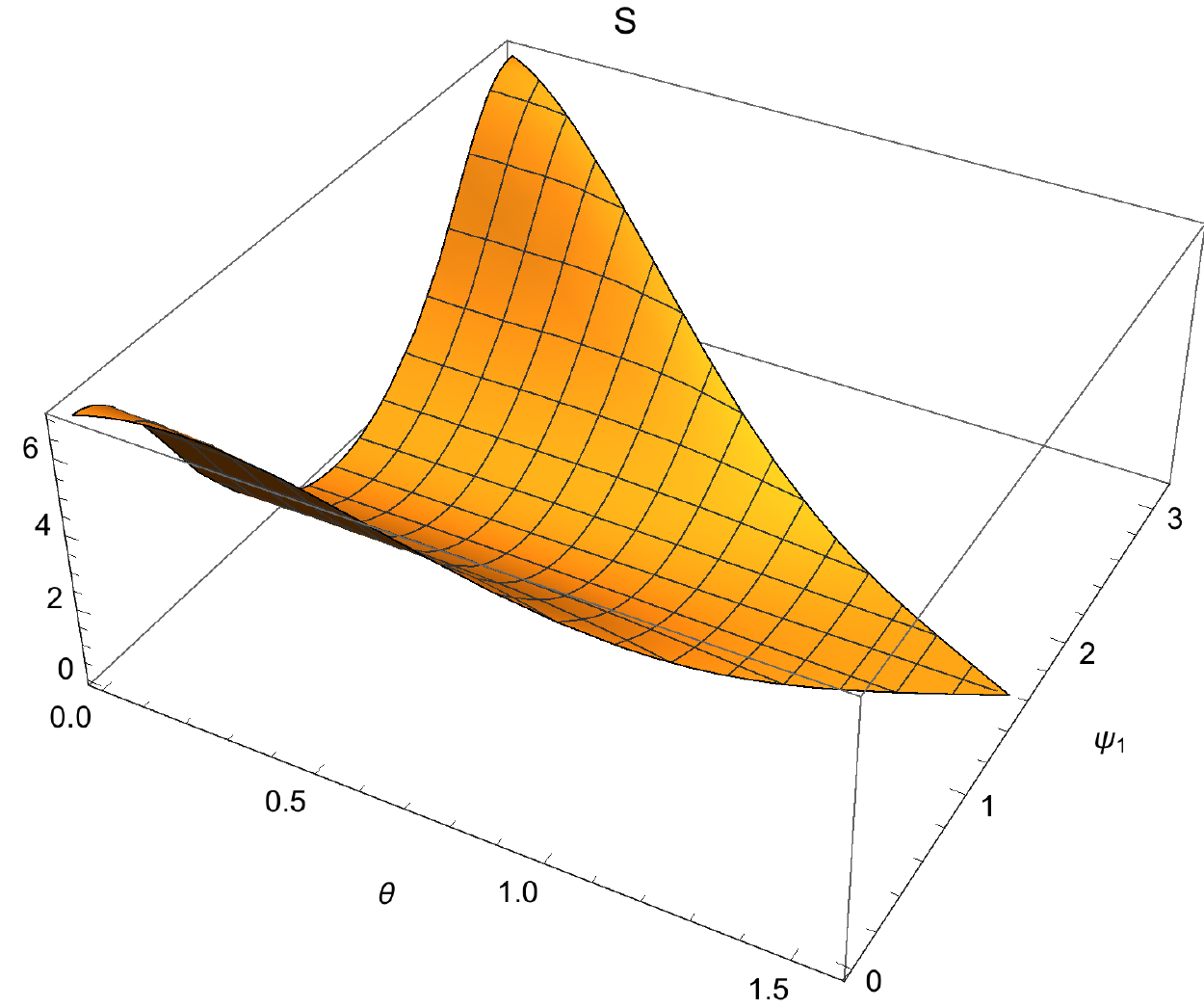}
\caption{The angular pattern $\text{S}$ in (\ref{fnl0}) in the allowed domain of $\theta$ and $\psi_1$. }
\label{angular2}
\end{figure}
\begin{figure}[t]
\includegraphics[angle=0, scale=0.8]{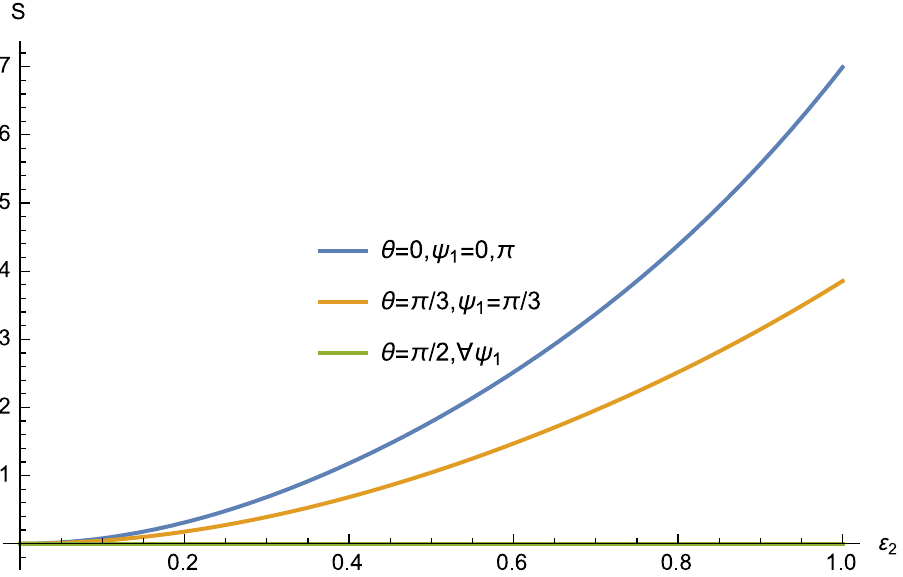}
\caption{The amplitude of the factor $\text{S}$ in (\ref{fnl0}) as a function of $-1<\vare_2<1$ for three angular configurations. The amplitude reaches its maximum at the poles, when the plane of the three momenta is (anti-)collinear with the preferred direction. For the configurations whose plane is perpendicular to the preferred direction the correction to the standard slow-roll result is zero.} 
\label{angular1}
\end{figure}

In this subsection we consider the bispectrum in the case $\chis=0$. We recall that with the phase $\phis=\pi/4$ there is then no constraint from the quadrupole piece of the two-point correlation, and the higher order pieces are under control as long as $\vare_2,\delta_2 < 1$. The backreaction bound can be satisfied given a low enough mass scale $M$ in (\ref{bound0}). To the first order in $\vare_2,\delta_2 < 1$, we obtain that, when $\chis=0$,
\be
f_{NL}^{(0)} = -\frac{10 \epsilon k_1}{3k_3}\lp c^2_{{\hat k}_1} + c^2_{{\hat k}_2}\rp\cos 2\phis \vare_2 + \mathcal{O}(\vare_2^2)\,.
\ee
For $\phis\neq \pi/4$, noting \eqref{B-anisotropic} and the bound exist on the quadrupole term, 
\be
-0.25\lesssim f_{\rm NL, max}^{(0)} \lesssim 0.28
\ee
which is small. Negative values of $f_{NL}^{(0)}$ is obtained for $\vare_2<0$. On the other hand, this contribution vanishes when $\phis=\pi/4$ and at that limit we are not restricted to tiny $\vare_2$ from the quadrupole constraint. Let us for simplicity fix the phase exactly to $\phis=\pi/4$ in all directions, so that also $\delta_2=0$. We then obtain the full result as
\be
%\label{fnl0}
f_{NL}^{(0)} = \frac{5 \epsilon k_1}{3k_3} \lb \frac{2 \sech_{2\times 3}}{\cosh_{2\times 1}+\cosh_{2\times 2}} \text{S}\rb\,,
\ee
where
\be
\text{S} = 2\sinh^2_{1+2}+\sinh_{2\times 3}\lp \sinh_{2\times 1} + \sinh_{2\times 2}\rp\,, \nonumber
\ee
and we have resorted to the short-hand notation such that
\beqa
\sech_{2\times 3} & = & \sech\lp 2\vare_2 c^2_{{\hat k}_3}\rp\,, \nonumber \\
\sinh_{1+2} & = & \sinh\lb \vare_2\lp c^2_{{\hat k}_1} + c^2_{{\hat k}_2}\rp\rb\,, \quad \text{etc}\,. \nonumber
\eeqa
If we set up our configuration the same way as in the previous case, we can write $c^2_{{\hat k}_1} + c^2_{{\hat k}_3} \approx  \cos^2\theta$
and $c^2_{{\hat k}_2} \approx c^2_{{\hat k}_1}$. We then have that
\be
\label{fnl0}
f_{\rm NL}^{(0)} = \frac{5 \epsilon k_1}{3k_3}\text{S}(\theta,c^2_{{\hat k}_1})
\ ,
\ee
with
\be
\text{S} =  2\tanh_{2\times 1}\sech_{2\times (\theta-1)}\lp\sinh_{2\times 1} +\sinh_{2\times (\theta-1)}\rp\,, \nonumber
\ee
where 
\beqa
\sinh_{2\times (\theta-1)} &\equiv&\sinh(2\cos(\theta)-2c_{k_1}^2)\,,\nonumber\\
\sech_{2\times (\theta-1)}&\equiv&\sech(2\cos(\theta)-2c_{k_1}^2)\,.\nonumber
\eeqa
The result (\ref{fnl0}) was written as the isotropic prediction (which is a number typically around $4-5$) times an angular modulation
function $\text{S}$. 
We plot the angular pattern of this function in the figure \ref{angular2}, where it becomes apparent that the {\it bipolar bispectrum} starts with a quadrupole rather than a monopole. The correction to the non-gaussianity is completely zero for the configurations whose planes are perpendicular to the preferred direction. In this case the amount of non-gaussianity is given by the slow-roll results. From the figure \ref{angular1}, where we numerically evaluate the amplitude of the function $\text{S}$, it is seen that the maximal $f_{\rm NL}^{(0)}$ could viably be of the order of about 20-30 for $\vare_2\lesssim 1$ for the coplanar configurations in which large momenta are parallel or anti-parallel with the preferred direction. This should indeed be detectable by near-future cosmological missions.  Thus, a smoking gun signature of the scenario is the extremely bipolar feature of the non-gaussianity. The nontrivial higher order statistics manifest only in preferred directions in the sky and disappear when correlating structures are in the perpendicular plane. As this can occur while the power spectrum remains the standard isotropic one, unexpected asymmetries could be awaiting their detection in the CMBR.
\section{Conclusions}
We considered non-Bunch Davies initial conditions for inflation that break rotational invariance. It was argued that the
4-parameter ansatz (\ref{beta-expansion}) should capture the essential features of various different scenarios. In addition there must be a scale $M$ such that the vacuum is not occupied for arbitrarily high modes, as that would lead to divergent energy density (and a too strong backreaction to the cosmological background before that). In (\ref{beta-expansion}), there are two parameters for the isotropic properties of the vacuum, $\varphi_S$ and $\chis$, which correspond to the phase difference of the Bogoliubov coefficients and their magnitude, respectively. Analogously, there are two parameters, $\delta_2$ and $\vare_2$ that quantify the possible directional dependence of the phase difference and the amplitude of particle creation, respectively, incorporating the cases where the primordial vacuum state might appear asymmetric to an observer in our frame.
\par
We distinguished two physically different classes of scenarios which might exhibit such asymmetry and focused on their predictions. They correspond to the following limits of the parameter $\chis$:  
\begin{itemize}
\item $\chis \gg 1$. Large $\chis$ means efficient particle creation. The backreaction constraints then push the phase $\varphi_S \rightarrow \pi/2$, and allow the scale $M$ to be an order-of-magnitude-or-so above the inflationary scale, so that the effective field theory approach could be reasonable. Then high energy processes above the scale $M>H$ could be responsible for the non-emptiness of the vacuum, and such processes could bring about anisotropic effects at suitable scales. 
\item $\chis = 0$. Then the vacuum is nontrivial purely due to the asymmetry. In this case, a special phase difference turns out to be $\varphi_S\rightarrow \pi/4$, since when the vacua are ''perpendicular'' to each other, there is no quadrupole in the power spectrum.
 If only a bubble not excessively larger than the observable universe was inflated from a smooth patch within a larger asymmetric region, the longest-wavelength pertubations in the universe could originate from an anisotropic vacuum. 
\end{itemize}
In both cases, the parameter $\vare_2$ typically determines the magnitude of observable effects, whilst the directional dependence of the phase difference, $\delta_2$, is more difficult to constrain.  

We found that in the case $\chis\gg 1$, the quadrupole constraint from the power spectrum still leaves room for non-Gaussian statistics of the three-point function. The $f_{NL}$ is amplified with respect to the standard case and has a subtle angular modulation. These are at the level that, optimistically, can be ruled out with the future data.

In the purely anisotropically modulated models with $\chis = 0$ the aforementioned features can be more pronounced. Firstly, the non-Gaussianity can be very significant, because the parameter $\vare_2$ does not necessarily produce a quadrupole $B$. Secondly, the angular modulation of the statistical properties is stronger, because there is no constant background $f_{NL}$ in contrast to the former class of models.

In this study we focused on the breaking of the spatial rotation symmetry. Of the Minkowski isometries, one could also consider constraining the breaking of the boost rotation symmetry, or of the translational invariance.  In the toy model \cite{Ashoorioon:2015pia} a particular such generalisation was realised by an $x$-dependent modulation of the initial condition, but a more proper analysis would require the development of a formalism, where the mode functions (when written in the conventional plane wave basis) are not orthogonal, but can be correlated \cite{Pereira:2015pga,Carroll:2008br}. In future we hope to extend the results of this paper by studying the inhomogeneous vacuum and the ensuing non-trivially correlated cosmology. 
\label{secC}
\acknowledgments
A.~A.~would like to thank Mahmoud Safari for discussions. 
\appendix
\section{Reality of position space spectrum} 
\label{AppA}
This appendix is written to provide further evidence for the generality of the vacua spanned by the parameter space (\ref{beta-expansion}) by showing that the initial conditions of the modes can not depend upon the handedness of their associated wave-vector (as otherwise their correlations in the position space would not be real).
 
In fact, we cannot produce imaginary terms in the primordial spectrum from our vacuum modification, because the modification is given by the real amplitude $\gamma$.
Nevertheless, suppose we generalised the parameterisation (\ref{beta-expansion2}) as 
\beqa \label{beta-expansion2}
\beta_{\vec k} & = & 
e^{-\frac{k}{aM}+i\phi}\sinh\chi\,, \\
  \chi & = &  \chis + \vare_1c_{\hat k} + \vare_2\,c_{\hat k}^2 + \vare_3\,c_{\hat k}^3\,, \\
  \phi & = & \phis + \delta_1c_{\hat k} + \delta_2\,c_{\hat k}^2 + \delta_3\,c_{\hat k}^3\,.
\eeqa 
In general, there would then appear a dipole,
\be
A = -2\frac{\vare_1\cos 2\phi\cosh 2\chi - \lp \vare_1+\delta_1\sin 2\chi\rp\sinh 2\chi}{\cosh 2\chi-\cos 2\phi\sinh 2\chi}\,.
\ee
For example, setting $\phi=\pi/2$, we have
\beqa
A& = & 2\vare_1, \quad B  =   2\lp\vare_1^2+\vare_2-\delta_1^2e^{-2\chi}\sinh 2\chi\rp\,, \nonumber \\
\frac{3}{2}C& = &  3e^{-4\chi}\lp\delta_2-\delta_1\vare_1\rp\delta_1  \nonumber \\
     & + & 2\vare_1^3 + 6\vare_2\vare_1
     - 3\delta_1^2\delta_2-3\delta_1^2\vare_1 \,.
 \eeqa

If the phase difference is $\phi=\pi/4$, we have
\beqa
A& = & 2\lp\delta_1+\vare_1\rp\tanh 2\chi, \nonumber \\
B & = & 2\vare_1\lp2\delta_1+\vare_1\rp + 2\lp \delta_2+\vare_2\rp\tanh 2\chi\,, \nonumber \\
\frac{3}{4}C & = & 3\delta_2\vare_1 + 3\lp\delta_1+\vare_1\rp \vare_2 \nonumber \\
& + & \lp \vare_1^3 + 3\delta_1\vare_1^2 -\delta_1^3 \rp\tanh 2\chi\,. \label{spectrum2}
 \eeqa
This is the special case where the power spectrum vanishes if we have a purely even-parity ($\delta_1=\delta_3=\vare_1=\vare_3=0$) anisotropic ($\chi=0$) distortion of the vacuum. 
Though we have written explicitly only the results only up to the third order in the starting point (\ref{beta-expansion2}) and the result (\ref{spectrum2}), the same conclusions would hold at any order. 

A question arises whether the we could have odd Bogoliubov coefficients without generating odd terms in the power spectrum. We will finally consider the case $\chi=0$,
\beqa
A& = & -2i\vare_1\cos 2\phi, \nonumber \\ 
B & = &   -2\lp\vare_2\cos 2\phi + \vare_1\lp \vare_1 - 2\delta_1\sin 2\phi\rp \rp\,, \nonumber \\
 C & = & 4\vare_1\vare_2 + \lp 4\delta_1^2\vare_1 - \frac{4}{3}\vare_1^3\rp\cos{2\phi} \nonumber \\
 & + & 4\lp\delta_2\vare_1 + \delta_1\vare_2\rp \sin{2\phi}\,.
 \eeqa
We can eliminate the dipole by choosing the difference $\phi=\pi/4$, the case in which also the quadrupole vanishes. Furthermore, by setting $\delta_1 = -\varepsilon_1/2$ we can eliminate the octopole. In fact, we could eliminate {\it any} contribution to the power spectrum at an arbitrary order. Taking into account higher order expansion of the initial condition, and requiring multipoles at successive orders to vanish, we are lead to the conditions 
\beqa
-2\delta_1 & = & \vare_1\,, \nonumber \\
-2\delta_2 & = & \vare_2\,, \nonumber \\
-12\delta_3 & = & \vare_1^3+6\vare_3\,, \nonumber \\
-4\delta_4 & = & \vare_1^2\vare_2+2\vare_4\,, \nonumber \\
-48\delta_5 & = & \vare_1^5-12\vare_1\vare_2^2+12\vare_1^2\vare_3+24\vare_5 \nonumber \\
 & \text{etc} & \,.
\eeqa
One could continue the process to an arbitrary order.
There are thus an infinite number of anisotropic distortions of the Bunch Davis vacua which have no consequences for the two-point correlations, and there are vacua with odd Bogoliubov coefficients but real power spectrum. However, one may conjecture that such properties could not be extended to the correlations of an arbitrary order in a non-trivial model.

We then show that odd powers of the angular parts of the power spectrum in momentum space have to be imaginary, otherwise the power spectrum in the position space will be imaginary. Noting that
\be\label{pwr-position}
\mathcal{P}_S(\mathbf{x})=\frac{1}{(2\pi)^{3/2}} \int d^3 {\mathbf{k}} ~\mathcal{P}_S(\mathbf{k}) \exp(i \mathbf{k}\cdot\mathbf{x}).
\ee
Aligning the preferred direction $\hat n$ along the $\hat z$, the integral over the azimuthal angular part of the above integral for the $n$-th multipole term in the primordial power spectrum is proportional to 
\be
\int_{0}^{\pi} \cos^n\theta\exp(i k x \cos(\theta))~d\cos\theta,
\ee 
which is 
\be
-i\frac{2  k \,
   _1F_2\left(\frac{n}{2}+1;\frac{3}{2},\frac{n}{2}+2;-\frac{k^2}{4}\right)}{n+2},
\ee
for odd $n$.  For even $n$, it is proportional to
\be
-\frac{2 \,
   _1F_2\left(\frac{n}{2}+\frac{1}{2};\frac{1}{2},\frac{n}{2}+\frac{3}{2};-\frac{k^2}{4}
   \right)}{n+1}
\ee 
where  $_1F_2$ is the generalized hypergeometric function and $k^2\equiv \mathbf{k}\cdot\mathbf{k}$. To compensate for the extra factor of $i$ in the coefficient of odd multipole terms, their coefficients should be odd too. 
\par
However, we see from (\ref{pwr-position}) that such imaginary pieces would imply that $\mathcal{P}_S(\mathbf{x}) \neq \mathcal{P}_S(-\mathbf{x})$. 
The real space $\mathcal{P}_S(\mathbf{x})$ quantifies correlations between structures separated by $\mathbf{x}$, something that can be meaningfully defined in a volume sufficiently bigger than $x^3$. Say we correlate structures $A$ with the structures $B$ at separated by $\mathbf{x}$. If $\mathcal{P}_S(\mathbf{x})$  is not $\mathcal{P}_S(-\mathbf{x})$ , we obtain another result by saying that we instead correlate the structures $B$ with the structures $A$. However, the two procedures of measuring the correlations are physically identical, and it is thus difficult to make sense of the distinction in an ordinary spacetime. In a non-commutative theory, one could accommodate such a feature, then stipulating that classical observations give the average result of the asymmetric correlation. Then, when one considers two-dimensional projections of the three-dimensional correlation spectrum, the left-right asymmetry becomes observable in a physically completely consistent way. Therefore the CMBR correlations would be well-defined in the presence of imaginary pieces in the primordial power spectrum, and they would include left-right asymmetric modulations, but that would require to allow for non-commutativity.  
\section{The backreacting energy density}
\label{AppC}
In this Appendix we will compute the backreacting energy density more accurately and taking into account the possible effects of anisotropies and nonzero mass of the modes. 
Instead of the sharp cut-off in~\eqref{beta0}, we will use here a smoothened exponential filter, 
\be
\label{beta0}
\beta_{\vec k}=e^{-\frac{k}{aM}}
\,\beta_0(\hat k)
\ .
\ee
We define the energy density residing in the excited modes (\ref{beta-expansion}) by extending the integral given by (\ref{enonbd})
over all scales 
\beqa
 \rho_{\text{non-BD}} & = & \frac{1}{a^3}
\int_0^{\infty} \frac{{\rm d}^3 k}{(2\pi)^3} \sqrt{m^2+\left(\frac{k}{a}\right)^2} |\B|^2 \nonumber \\
& = & \frac{M^4}{4\pi^2}\sinh^2\lp\chis\rp f_{\frac{m}{M}} f_{\vare_2}\,. 
\eeqa
The proportionality to the occupation number and the fourth power of the cut-off is 
modulated by the functions $f_{\frac{m}{M}}$ and $f_{\vare_2}$ due to mass of the modes and
anisotropy, respectively. 
The former differs from unity as
\be 
f_{\frac{m}{M}} \eqsim 1+\frac{m^2}{M^2}\quad m \lesssim M\,.
\ee
The piece $f_{\vare_2}$ is more subtle and depends also on $\chis$. 
In the limit of large isotropic occupation number, it becomes 
\be
\lim_{\chis \rightarrow \infty} f_{\vare_2} = \frac{1}{2}\sqrt{\frac{\pi}{2\vare_2}}\text{Erfi}\lp \sqrt{2\vare_2}\rp\,.
\ee
In the opposite limit $f_{\vare_2}$ diverges quadratically and thus a purely anisotropic contribution does not cancel out
\be
\lim_{\chis \rightarrow 0} \rho_{\text{non-BD}} = \frac{M^4}{20\pi^2}f_{\frac{m}{M}}\lp \vare_2^2 + \frac{5}{27}\vare_2^4 + \dots \rp\,.
\ee
In typical inflationary scenarios $m \ll H$, and thus for the sub-horizon modes $m\rightarrow 0$ gives a good approximation for the dispersion relation. Also, the anisotropy
of the non-standard vacuum also typically affects the result only by a factor of the order of unity.

The constraint (\ref{background-backreaction})  together with the result for the modified power spectrum implies the hierarchy 
\be \label{bound}
\lp \frac{M}{M_{\text{Pl}}}\rp^4 \lesssim \frac{32\pi^2}{\gamma_S\sinh^2\chis f_{\frac{m}{M}} f_{\vare_2}}\lb\epsilon \eta {\mathcal P}_S\rb_0\,,
\ee
where the square brackets is determined from observations\footnote{The $\epsilon$ and $\delta$ can depend somewhat on the model even given the observed spectral dependence.}. Written relative to the Hubble parameter, the upper bound on the cut-off scale is 
\be
\lp \frac{M}{H}\rp^4 \lesssim \frac{\gamma_S}{2\sinh^2\chis f_{\frac{m}{M}}f_{\vare_2}} \lb \frac{\eta}{{\mathcal P}_S}\rb_0\,.
\ee 
We can then look at what these bounds imply in the limits of large and small isotropic amplitudes $\chis$.

Since with large occupation numbers we have that
\be
\lim_{\chis \rightarrow \infty} \frac{\gamma_S}{2\sinh^2\chis} = 1-\cos{2\phis}\,,
\ee
the upper bound is maximal with the phase difference $\phis=\pi/2$. However, we note that because in this limit
\be
\lim_{\chis \rightarrow \infty} \frac{1}{2\gamma_S \sinh^2\chis} = 0\,,
\ee
the bound (\ref{bound}) restricts us to consider ultra-low scale inflation with increasing $\chis$. Thus, the case 
$\chis>1$ together $M \gtrsim H$ requires very delicate balancing of scales.  On the other hand, in the limit of small occupation number we have the restriction
\be \label{bound0}
\vare_2^2 \lesssim \frac{160\pi^4}{f_{\frac{m}{M}}}\lp\frac{M_{\text{Pl}}}{M}\rp^4 \lb \epsilon^2\eta{\mathcal P}_S\rb_0\,, \, \text{when} \, \chis=0\,.
\ee
Thus, if we want to consider larger values of the $\vare_2$, we need to adjust the cut-off scale $M$ to a lower value.

\section{Coefficients of quadratic terms} \label{AppB}

The coefficients of $\xi_{lm;l'm'}^{(2)}$, in eq. \eqref{xiexpansion} could be computed in terms of $l$ and $m$'s
\begin{eqnarray}
&&\xi_{lm;l'm'}^{--}=-\delta_{m',m+2}\times\nn
 &&\left[\delta_{l',l}{\sqrt{
(l^2-(m+1)^2)(l+m+2)(l-m)}\over (2l+3)(2l-1)}\right.\nonumber\\
&&\left.-{1 \over
2}\delta_{l',l+2}\sqrt{{(l+m+1)(l+m+2)(l+m+3)(l+m+4)\over
(2l+1)(2l+3)^2(2l+5)}}\right. \nonumber \\ 
&& \left. -{1 \over 2}\delta_{l',l-2}{\sqrt{ {(l-m)(l-m-1)(l-m-2)(l-m-3)\over (2l+1)(2l-1)^2(2l-3)}}}\right] ,\nonumber
\end{eqnarray}
\begin{eqnarray}
&&\xi_{lm;l'm'}^{++}=\xi_{l'm';lm}^{--},\nonumber
\end{eqnarray}
\begin{eqnarray}
&&\xi_{lm;l'm'}^{+-}={1 \over 2}\delta_{m',m}\left[-2 \, \delta_{l',l}
\frac{(-1+l+l^2+m^2)}{(2l-1)(2l+3)} \right. +\nn
&&\left. \delta_{l',l+2}\sqrt{\frac{((l+1)^2-m^2)((l+2)^2-m^2)}{(2l+1)(2l+3)^2(2l+5)}}\right.
\nonumber \\
&& +\left.\delta_{l',l-2}\sqrt{\frac{(l^2-m^2)((l-1)^2-m^2)}{(2l-3)(2l-1)^2(2l+1)}}\right],
\nonumber
\end{eqnarray}

\begin{eqnarray}
&&\xi_{lm;l'm'}^{-0}=-\frac{1}{\sqrt{2}}\delta_{m',m+1}\left[
\delta_{l',l} \frac{(2m+1)\sqrt{(l+m+1)(l-m)}}{(2l-1)(2l+3)} \right. \nn
&&\left. +\delta_{l',l+2}\sqrt{\frac{((l+1)^2-m^2)(l+m+2)(l+m+3)}{(2l+1)(2l+3)^2(2l+5)}}\right.
\nonumber \\
&& \left. -
\delta_{l',l-2}\sqrt{\frac{(l^2-m^2)(l-m-1)(l-m-2)}{(2l-3)(2l-1)^2(2l+1)}}
\right], \nonumber
\end{eqnarray}

\begin{eqnarray}
&&\xi_{lm;l'm'}^{+0}=-\xi_{l'm';lm}^{-0},\nonumber 
\end{eqnarray}

\begin{eqnarray}
&&\xi_{lm;l'm'}^{00}=\delta_{m,m'} \left[ \delta_{l,l'} \frac{(2l^2+2l-2m^2-1)}{(2l-1)(2l+3)} +\right. \nn
&&\left. \delta_{l',l+2}
\sqrt{\frac{((l+1)^2-m^2)((l+2)^2-m^2)}{(2l+1)(2l+3)^2(2l+5)}}
\right. \nonumber \\ 
&&\left. +\delta_{l',l-2} \sqrt{\frac{(l^2-m^2)((l-1)^2-m^2)}{(2l-3)(2l-1)^2(2l+1))}} \right].\nonumber
\end{eqnarray}

\end{document}